\documentclass{nature}

\usepackage{graphicx}
\usepackage{float}
\usepackage{amsmath}
\usepackage{amssymb}
\usepackage{bibentry}
\usepackage{amsfonts}
\usepackage{array}
\usepackage{listings}
\usepackage{lmodern}
\usepackage{mathpazo}
\usepackage{microtype}
\usepackage{hyperref}
\usepackage{xfrac}
\usepackage{color}
\usepackage[version=3]{mhchem} 
\usepackage{flushend}
\usepackage{siunitx}
\usepackage{tabularx}
\usepackage[font=small,labelfont=bf]{caption}
\usepackage{graphicx}
\makeatletter
\let\saved@includegraphics\includegraphics
\AtBeginDocument{\let\includegraphics\saved@includegraphics}
\renewenvironment*{figure}{\@float{figure}}{\end@float}
\makeatother

\DeclareSIUnit\bar{bar}
\usepackage{changes}

\author{Outhmane Chahib,$^{1}$ Yulin Yin,$^{2}$ Jung-Ching Liu,$^{1}$ Chao Li,$^{1}$ Thilo Glatzel,$^{1}$ Feng Ding,$^{2}$ Qinghong Yuan,$^{3}$  Ernst Meyer$^{1,*}$ \& Rémy Pawlak$^{1,*}$}

\title{Probing charge redistribution at the interface of self-assembled \textit{cyclo}-$P_{\rm 5}$ pentamers on Ag(111)}

\begin{document}
\maketitle
\begin{affiliations}
\item Department of Physics, University of Basel, Klingelbergstrasse 82, 4056 Basel, Switzerland
\item Faculty of Materials Science and Engineering/Institute of Technology for Carbon Neutrality, Shenzhen Institute of Advanced Technology, Chinese Academy of Sciences, Shenzhen 518055, China
\item State Key Laboratory of Precision Spectroscopy School of Physics and Electronic Science, East China Normal University, 500 Dongchuan Road, Shanghai 200241, China
 \end{affiliations}
\section*{\large \textsf{Abstract}}
Phosphorus pentamer (\textit{cyclo}-$P_{\rm 5}^-$) ions are unstable in nature but can be synthesized at the Ag(111) surface. Unlike monolayer black phosphorous, little is known about their electronic properties when in contact with metal electrodes, although this is crucial for future applications. Here we characterize the atomic structure of \textit{cyclo}-$P_{\rm 5}$ assembled on Ag(111) using atomic force microscopy with functionalized tips and density functional theory. Combining force and tunneling spectroscopy, we find that a strong charge transfer induces an inward dipole moment at the \textit{cyclo}-$P_{\rm 5}$/Ag interface as well as the formation of an interface state. We probe the image potential states by field-effect resonant tunneling and quantify the increase of the local change of work function of 0.46 eV at the \textit{cyclo}-$P_{\rm 5}$ assembly. Our results suggest that the high-quality of the \textit{cyclo}-$P_{\rm 5}$/Ag interface might serve as a prototypical system for electric contacts in phosphorus-based semiconductor devices.

\paragraph{Keywords:} \textit{cyclo}-$P_{\rm 5}^-$ pentamer, work function, atomic force microscopy, scanning tunneling microscopy, ﬁeld-emission resonance spectroscopy, density functional theory

\section*{\large \textsf{Introduction}}
Elemental phosphorus (P) is not only ubiquitous in human life, it is also one of the most fascinating areas of chemistry as it can exist in a large diversity of allotropes,\cite{Liu2015,Carvalho2016,Batmunkh2016} in various cluster configurations \cite{Jones1990,Chen2000} or in organic compounds.\cite{Giusti2021} The phosphorous polymorphism is even multiplied on the atomic-scale when using a surface to constrain the reaction in two dimensions (2D). As in the field of on-surface chemistry producing complex nanographene structures in ultra-high vacuum (UHV),\cite{Cai2010,Clair2019} surface-assisted phosphorus reactions on metals have synthesized blue phosphorus,\cite{Zhang2016} P chains\cite{Zhang2021} or even planar \textit{cyclo}-$P_{\rm 5}$ rings.\cite{Zhang2020,Yin2022} Since then, phosphorus allotropes have emerged as a promising one-atom thick 2D material beyond graphene, due to its moderate direct band gap (0.3 to 2.0 eV)\cite{Li2017} suitable for nanoelectronics and nanophotonics applications.\cite{Liu2014,Wang2015} However, allotropic configurations, their atomic buckling, defects or potential alloy formation can be detrimental for the semiconducting character. In addition, the interaction of 2D materials with delocalized electrons of a metal, as well as the dynamical charge transfer between the two media, are key factors in fostering new gate-tunable functionalities such as superconductivity.\cite{Shao2014,Zhang2017}  Experimental study of these aspects at the fundamental level is therefore essential for future quantum applications where metallic electrical contacts are required.\cite{Li2014a} \\

\noindent
Low-temperature scanning probe microscopy is an incontrovertible tool for assessing  atomic structures in contact to metals and characterizing their electronic properties with high spectral resolution in UHV. Atomic force microscopy (AFM) with functionalized tips\cite{Gross2009,Moenig2018} has opened new avenues into the real-space imaging with improved lateral and vertical resolution of aromatic molecules, cyclo-carbon\cite{Kaiser2019} and monoelemental 2D materials,\cite{Liu2019,Pawlak2020} while charge distributions and work function changes at the nanometer scale are also accessible using Kelvin probe force microscopy (KPFM).\cite{Mohn2012,Schuler2014,Meier2017,Pawlak2017} The investigation of the local density of states (LDOS) of 2D materials near the Fermi level is readily achieved by means of scanning tunneling microscopy and spectroscopy (STM/STS). Tunneling spectroscopy can also probe the IPS of 2D synthetic materials such as graphene,\cite{Borca2010} germanene\cite{Borca2020} or borophene.\cite{Liu2021,Liu2022} Quantifying these Stark-shifted unoccupied states lying below the vacuum level give not only access to the fundamental physical processes involved in charge carrier dynamics but also to quantify local modulations of the work function at the interface between 2D materials and metals.

\noindent 
By applying this {\it in-situ} methodology, we determine here the structure of phosphorus chains and self-assembled \textit{cyclo}-$P_{\rm 5}$ pentamers on Ag(111) using low-temperature (4.5 K) AFM imaging with CO-terminated tips. KPFM spectroscopic measurements  indicates the formation of an inwards dipole moment at the $P_{\rm 5}$/Ag interface, which results from the charge transfer from the Ag substrate to the network as confirmed by DFT calculations. This charge transfer leads to a complex charge redistribution and the formation of an interfacial hybridized state (IS). Through field-emission resonance tunneling (FERT) and STS spectroscopy, we determined the energy position of the IS state and the series of IPS at the \textit{cyclo}-$P_{\rm 5}$ assembly as compared to pristine Ag, confirming an increase of the local work function of $\sim$ 0.46 eV. Given the strong interest in tailoring the physical characteristics of monoelemental 2D materials contacted to a metal, we think that the \textit{cyclo}-$P_{\rm 5}$/Ag interface might serve as a model system for future devices involving electric contacts. 

\section*{\large \textsf{Atomic-scale imaging of phosphorus chains and \textit{cyclo}-$P_{\rm 5}$ pentamers}}
\begin{figure}[t!]
		\centering
		\includegraphics[width=\textwidth]{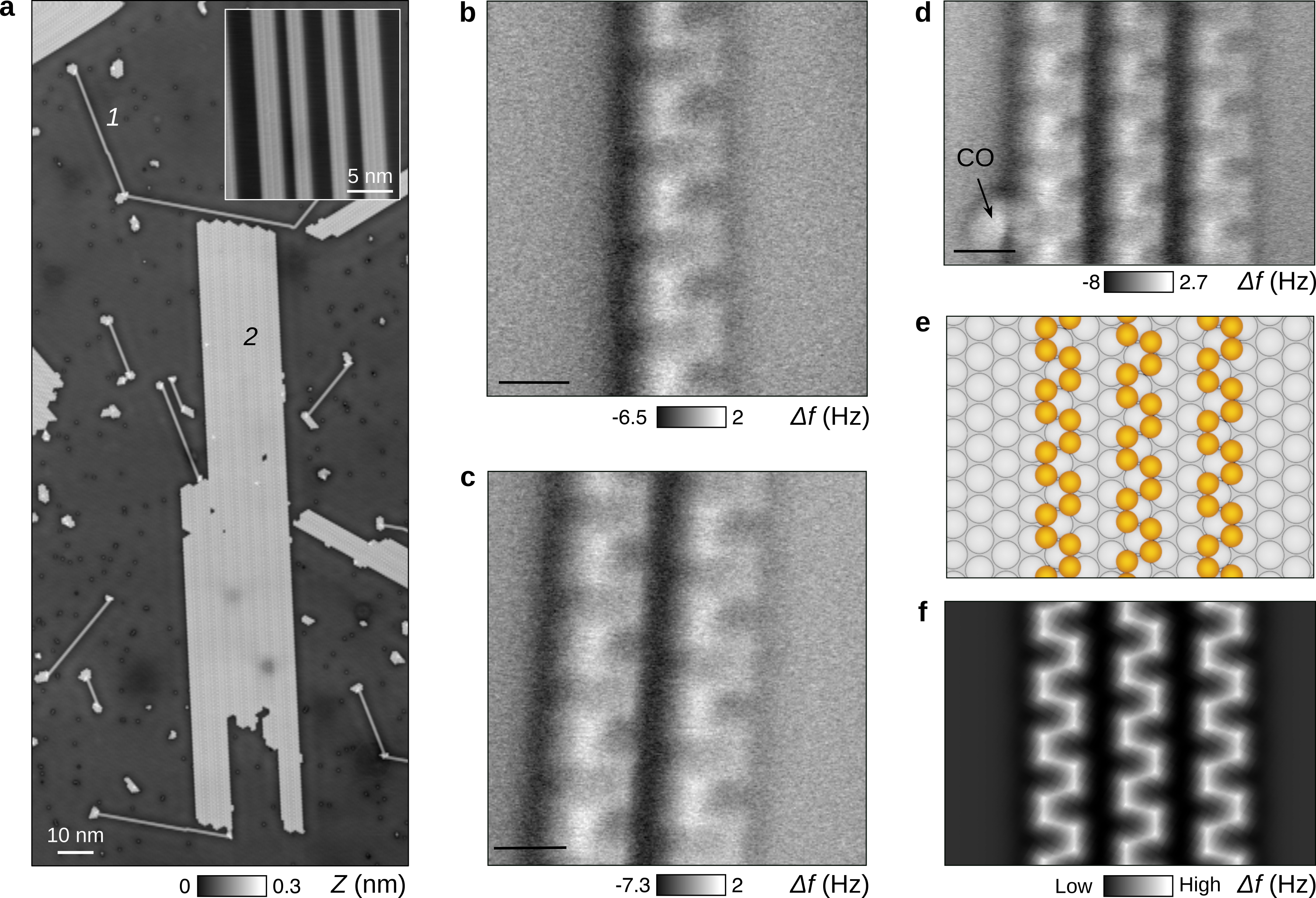}
		\caption{{\bf Atomic structure of phosphorus chains on Ag(111).}
		{\bf a}, STM topographic image after sublimation of phosphorus atoms on Ag(111) leading to P chains ({\it 1}) and \textit{cyclo}-$P_{\rm 5}$ domains ({\it 2}), ($I_{\rm T}$ = 1 pA, $V$ = 0.15 mV). The inset shows a STM image of the single, double and triple chains, respectively.  
		{\bf b-d}, Series of AFM images with CO-terminated tip revealing the armchair structure of single, double and triple P chains, ($f_{\rm 0}$ = 26 kHz, $A$ = 50 pm). Scale bars are 1 nm.  
		{\bf e}, Atomic configurations of the triple armchair chains obtained by DFT calculations. Phosphorus and silver atoms are shown in orang and gray, respectively.
		{\bf f}, Corresponding AFM simulation using the DFT coordinates. \label{Fig1}}
	\end{figure}

Phosphorus atoms were sublimed in UHV onto the Ag(111) substrate kept at about 150 °C (see Methods). Figure~\ref{Fig1}a shows an STM overview image  of the resulting structures for a relative coverage of less than 0.3 monolayer (ML). Extended 1D chains aligned along the $[1\bar{1}0]$ directions of Ag(111) (marked {\it 1} in Fig.~\ref{Fig1}a) coexist with domains of $P_{\rm 5}$ molecules ({\it 2}), as recently reported in references.\cite{Zhang2020,Zhang2021} The inset of Fig.~\ref{Fig1}a further shows a STM image of the double and triple chains, that depends on the P deposition rate.\cite{Zhang2020} Each chain configuration has a relative STM height of 1.6 \AA, and a width of 11 \AA, 17 \AA, and 26 \AA~for the single, double and triple chains, respectively. 

\noindent
To precisely determine their atomic configurations, we employed AFM imaging with CO-terminated tips (see Methods, Figs.~\ref{Fig1}c-d).\cite{Gross2009} A common AFM contrast is observed for all configurations assigned to an armchair structure of the chain, which resembles that of hydrocarbon chains.\cite{Giovanelli2022} The relaxed structure of the triple chain configuration calculated from DFT is shown in Fig.~\ref{Fig1}e. Phosphorus atoms colored in orange sit on bridge sites of the Ag lattice (gray) and are aligned along one $[1\bar{1}0]$ direction in accordance with the experimental data.  Based on the DFT coordinates we simulated the AFM image (see Methods, Fig.\ref{Fig1}f). The excellent agreement with the experimental image of Fig.~\ref{Fig1}d confirms the armchair structure of the P chains on Ag(111).\\

\begin{figure}[t!]
		\centering
		\includegraphics[width=\textwidth]{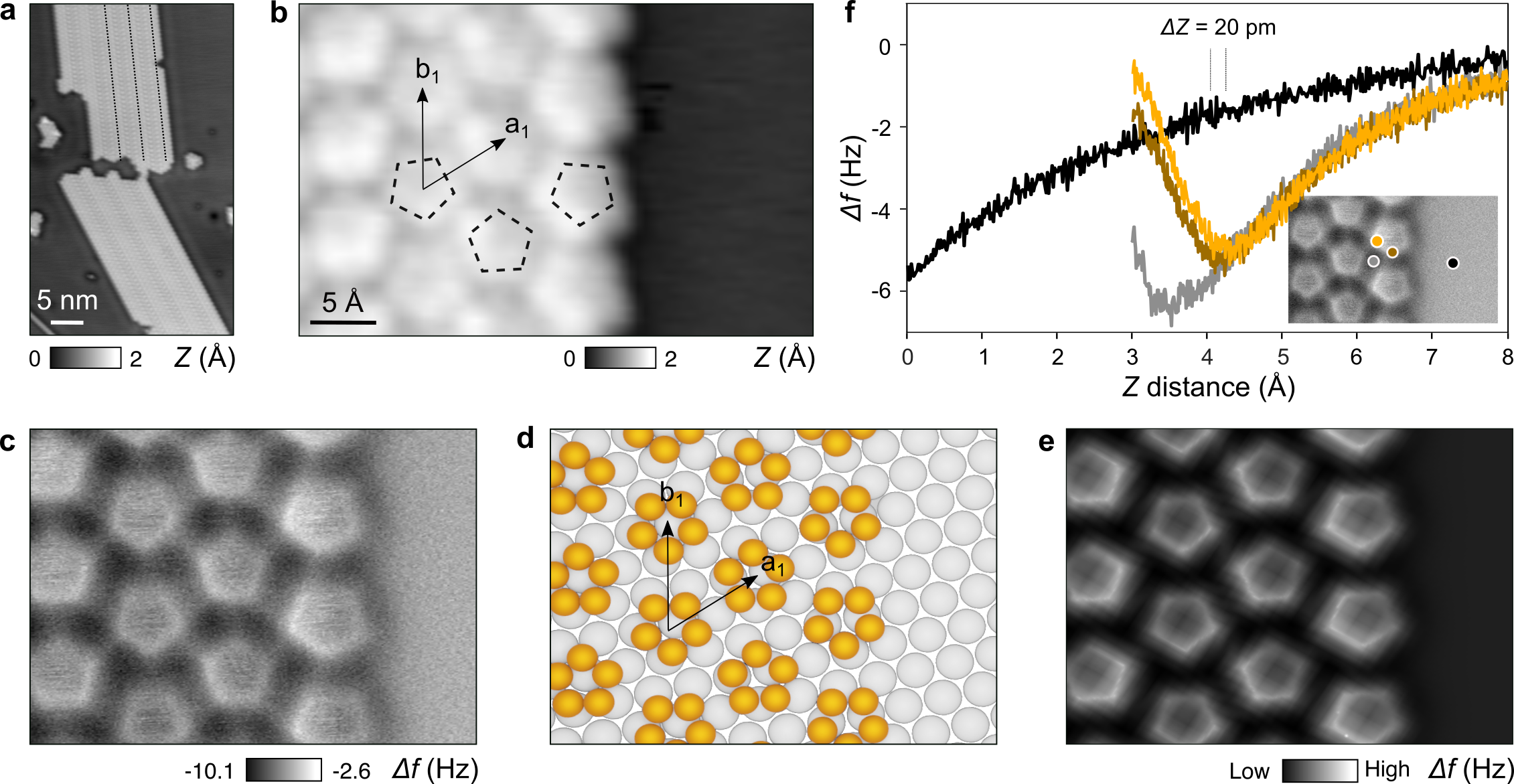}
		\caption{{\textbf{Atomic structure of the self-assembled \textit{cyclo}-$P_{\rm 5}$ molecules.}} 
		{\bf a}, STM image of the self-assembled pentamers on Ag(111), ($I_{\rm T}$ = 1 pA, $V$ = 0.15 mV). Islands systematically shows a superlattice of bright lines rotated by 19° with respect to the $[1\bar{1}0]$ directions of Ag(111). 		
		{\bf b}, Close-up STM topography showing the $P_{\rm 5}$ molecules depicted by dashed pentagons.  
		{\bf c}, Corresponding AFM image revealing the $P_{\rm 5}$ chemical structure, ($f_{\rm 0}$ = 26 kHz, $A$ = 50 pm). 
		{\bf d}, Atomic configurations of the pentamer assembly on Ag(111) obtained by DFT. Phosphorus and silver atoms are shown in orange and gray, respectively.
		{\bf e}, Corresponding AFM simulation using the DFT coordinates.
		{\bf f}, Site-dependent $\Delta f(Z)$ spectroscopic curves acquired at one P atoms of a $P_{\rm 5}$ molecule (orange), between two $P_{\rm 5}$ molecules (brown) and on Ag(111) (black), respectively. The local minima of the $\Delta f(Z)$ curves indicate the relative height of the phosphorus atoms.\label{Fig2}}
	\end{figure}

\noindent 
Increasing the P coverage to about 0.4 ML while keeping the substrate at 150°C leads to the formation of large islands of \textit{cyclo}-$P_{\rm 5}$ pentamers relative to the chains (Fig.~\ref{Fig2}a).\cite{Yin2022} In Fig.~\ref{Fig2}b, the close-up STM image reveals the structure of the self-assembled domains consisting of an hexagonal lattice with parameters $a_{1}$ = $b_{1}$ = 7.3 \AA~. Each bright protrusion corresponds to one \textit{cyclo}-$P_{\rm 5}$ molecule as schematized by the black dashed pentagons. Domains of \textit{cyclo}-$P_{\rm 5}$ rings also exhibit a superstructure characterized by stripes separated by $\approx$ 3.8 nm (i.e. 6 $P_{\rm 5}$ rows) as shown by black dotted lines in Fig.~\ref{Fig2}a. These lines are rotated by 19° as compared to the $[1\bar{1}0]$ directions of the Ag(111) substrate, which agrees with previous experimental works\cite{Zhang2020} as well as the relaxed structure obtained by DFT calculations (Fig.~\ref{Fig2}d).\cite{Yin2022} A deeper insights into the chemical structure of the \textit{cyclo}-$P_{\rm 5}$ molecules is provided by the AFM image of Fig.~\ref{Fig2}c. The P-P bond length within the pentagon extracted by AFM is about 2.2 \AA~, which is comparable to the value of 2.185 \AA~obtained by DFT. For comparison, we also simulated the AFM image based on the DFT coordinates, allowing us to confirm the exact position and structure of the  $P_{\rm 5}$ molecules in their self-assembly in registry with the Ag(111).\\ 

\noindent
To accurately quantify the atomic corrugation within the \textit{cyclo}-$P_{\rm 5}$ structure, we acquired a series of site-dependent $\Delta f(Z)$ spectroscopic curves (Fig.~\ref{Fig2}f), at the locations marked in the inset AFM image.  The black and gray curves were obtained on Ag and between two pentamers, respectively.  On top of neighboring atoms of a \textit{cyclo}-$P_{\rm 5}$ (orange and brown curve), the spectra exhibits a characteristic dip arising from the interaction between the front-end oxygen atom of the CO-terminated tip with the phosphorus atom. The dashed vertical lines indicate the $Z$ position of their bottoms and is the signature of the relative atomic  $Z$ height.\cite{Pawlak2020} The difference $\Delta Z$ of $\approx$ 20-30 pm thus represents the intrinsic atomic corrugation within the \textit{cyclo}-$P_{\rm 5}$ pentagonal structure,\cite{Chen2000} which is comparable with atomic corrugations in graphene\cite{Boneschanscher2014} or planar molecules.\cite{Kawai2018} 
Thus, this confirms the planarity of the \textit{cyclo}-$P_{\rm 5}$ structure,\cite{Chen2000} as reflected in the constant-height AFM image of Fig.~\ref{Fig2}e. 

\section*{\large \textsf{Charge distribution at the \textit{cyclo}-$P_{\rm 5}$/Ag interface}}
\begin{figure}[hbt!]
		\centering
		\includegraphics[width=0.65\columnwidth]{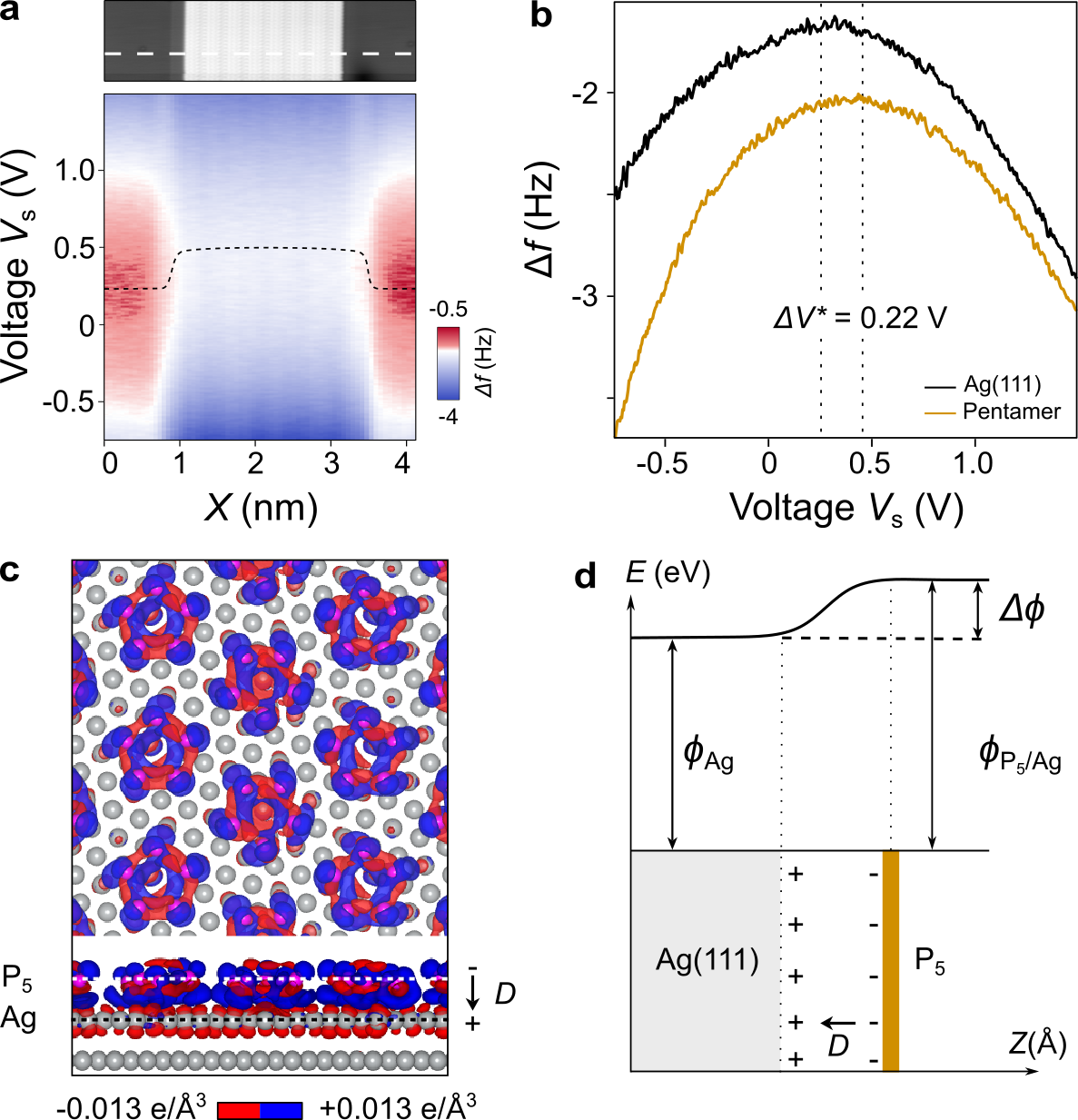}
		\caption{{\bf Charge redistribution at the \textit{cyclo}-$P_{\rm 5}$/Ag(111) interface.} 
		 {\bf a}, Frequency shift $\Delta f$ as a function of sample bias voltage $V_{\rm s}$, measured across a pentamer domain shown in the STM image (top), (parameters : $f_{\rm 0}$ = 26 kHz, $A$ = 80 pm).
		 {\bf b}, Single $\Delta f(V)$ curves at the pentamer assembly (orange) as compared to the Ag(111) (black). Dashed lines mark the top of the parabola allowing to extract a LCPD shift $\Delta V^*$ = 0.22 V. 		
		 {\bf c}, Top and side views of the charge redistribution between pentamers and Ag(111).  Blue areas show electron accumulation, red areas electron depletion. The isosurface level of the plot is set to  $\pm$ 13$\times$10$^{-3}$ e/\AA$^{3}$. 
		 {\bf d}, Schematic illustration of the charge redistribution at the $P_{\rm 5}$/Ag(111) interface leading to an inward surface dipole ($D$) moment and a local work function change ($\phi_{\rm P_5/Ag}$). The  \textit{cyclo}-$P_{\rm 5}$ layer is colored in orange.  $\Delta V^*$ refers to the LCPD change.
		\label{Fig3-1}}
	\end{figure}

The binding energy of $P_5$ pentamers has been calculated by DFT to be strong on Ag(111), allowing the stabilization of the \textit{cyclo}-$P_{\rm 5}$ structure through a charge transfer.\cite{Yin2022}  To provide insights into the charge distribution at the \textit{cyclo}-$P_{\rm 5}$/Ag interface, we performed force \textit{versus} voltage spectroscopic measurements (see Methods). Experimentally, the frequency shift $\Delta f$ as a function of the sample bias $V_{\rm s}$ is measured at a constant tip height $Z$, providing in the $\Delta f (V)$ curve a parabola due to the electric force acting between tip and sample. The voltage $V^*$ at top of the parabola represents the local contact potential difference (LCPD) between tip and sample, which allows one to image charge distributions and work function changes with nanoscale resolution.\cite{Mohn2012,Schuler2014,Meier2017,Pawlak2017} Figure~\ref{Fig3-1}a shows a $\Delta f (V)$ cross-section acquired across a $P_{\rm 5}$ domain (see STM inset of Fig.~\ref{Fig3-1}a). Single $\Delta f (V)$ point-spectra on top of the $P_{\rm 5}$ network (orange) and on Ag(111) (black ) are plotted in Fig.~\ref{Fig3-1}b, respectively. Dashed lines in both figures refers to the $V^*$ position. The LCPD value systematically shifts towards positive values ($\Delta V^*$ $\approx$ 0.22 V) for the pentamer assembly as compared to the pristine Ag substrate.  This indicates the accumulation of charges at the $P_{\rm 5}$ network as compared to the Ag substrate.

\noindent
To better rationalize this, we calculated the charge redistribution at the \textit{cyclo}-$P_{\rm 5}$/Ag interface (see Methods), whose top and side views of isosurfaces of electron accumulation (blue, +13$\times$10$^{-3}$ e/\AA$^{3}$) and depletion (red, -13$\times$10$^{-3}$ e/\AA$^{3}$) are displayed in Fig.~\ref{Fig3-1}c. An electron transfer from the Ag(111) substrate to the P atoms of the pentamers ( is observed as a charge accumulation located at the \textit{cyclo}-$P_{\rm 5}$ ring (red). In the $P_{5}$/Ag gap (marked by white and black dashed lines in the side view of Fig.~\ref{Fig3-1}c), charge accumulation/depletion layers emerges below each \textit{cyclo}-$P_{\rm 5}$ structure, which supports the formation of an hybridized state.\cite{Liu2021,Liu2022} We emphasize that such an interface state is not restricted to 2D Xenes on metals since it is well-established in organic/metal systems.\cite{GonzalezLakunza2008}

\noindent
Between \textit{cyclo}-$P_{\rm 5}$ rings, we note the absence of in-plane charge redistribution. Considering that the last Ag layer is depleted (red) while each \textit{cyclo}-$P_{\rm 5}$ has an excess of charges (blue), the $P_5$ assembly can be approximated to a lattice of surface dipole moments of $D$ = 1.42 Debye pointing towards the substrate (see arrow in Fig.~\ref{Fig3-1}c). This observation is consistent with an increase of the LWF at the $P_{5}$/Ag interface induced by an inwards dipole moment as schematized in Fig.~\ref{Fig3-1}d, which is in agreement with the increase of the LCPD in force spectroscopy. It is important to mention here that the LCPD value has a strong distance-dependence on metal substrate, which is good indicate of the local work function changes at the atomic scale, but prevents quantitative determination.\cite{Nony2009,Sadeghi2012} Indeed, the $\Delta V^*$ cannot directly account for the difference of work function $\Delta \phi$ = $\phi_{P_5/Ag}$ - $\phi_{Ag}$ shown in Fig.~\ref{Fig3-1}d, due to averaging effects of the electrostatic interactions between tip and sample. Last, the presence of an interfacial state can alter the amount of charge transfer away from an integer number compared to those with weaker interactions or adsorbed on insulating layers. Indeed, the Bader charge analysis show an accumulation of electrons on P atoms (-0.115 e) and an electron depletion  (+0.057 e) of the depleted Ag layer. Thus, we conclude that the \textit{cyclo}-$P_{\rm 5}$ does not have a pure anionic character for the $P_5$ molecule when  adsorbed on Ag(111) (i.e. \textit{cyclo}-$P_{\rm 5}^-$), as expected by theory for its gas-phase counterpart. \\

\section*{\large \textsf{Interface state and work function of the \textit{cyclo}-$P_{\rm 5}$ assembly}}
\begin{figure}[hbt!]
		\centering
		\includegraphics[width=\textwidth]{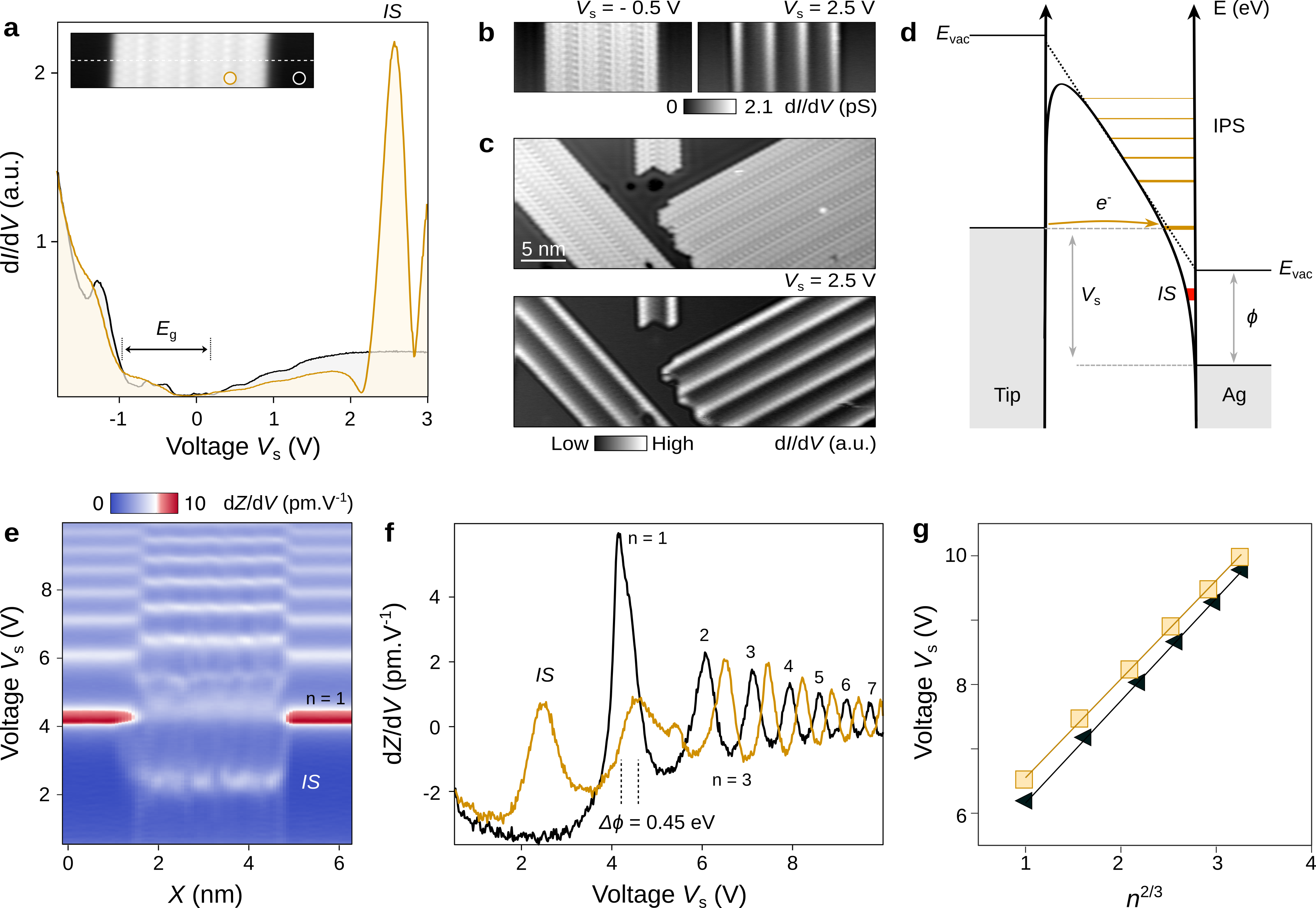}
		\caption{{\bf  Tunneling spectroscopy of the $P_{\rm 5}$/Ag interface.} 
		 {\bf a}, d{\it I}/d{\it V} point-spectra acquired above the $P_{\rm 5}$ assembly (orange) and on Ag(111) (black), where precise locations are shown in the STM inset.  (parameters: $I_{\rm t}$ = 1 pA, $V_{\rm s}$ = 500 mV, $A_{\rm mod}$ = 10 mV, $f$ = 511 Hz).
		 {\bf b}, d{\it I}/d{\it V} maps at $V_{\rm s}$ = -1.25 and 2.5 V corresponding to the valence band energy and the IS interface state , respectively. 
		 {\bf c}, STM topographic image of three $P_{\rm 5}$ domains and the corresponding  d{\it I}/d{\it V} maps of the IS modulation.
		 {\bf d}, Scheme of the band alignment and the formation of Stark-shifted IPS (orange lines).
		 {\bf e}, Field-effect resonance tunneling (FERT cross-section acquired across the $P_{\rm 5}$ assembly along the dashed line in {\bf a}, (Set-points: $I_{\rm t}$ = 1 pA, $V_{\rm s}$ = 500 mV, $A_{\rm mod}$ = 35 mV, $f$ = 511 Hz).  
		 {\bf f}, Single FERT spectra of the $P_{\rm 5}$ assembly and the Ag(111) substrate, showing the series of n$^{th}$ IPS.
		 {\bf g}, Extracted IPS peak voltages as a function of $n^{\rm 2/3}$.  \label{Fig3}}
	\end{figure}

To shed more lights into the electronic properties at the $P_{\rm 5}$/Ag interface, we next performed differential conductance measurements (d\textit{I}/d\textit{V}) across one $P_5$ domain (see Methods). Figure~\ref{Fig3}a shows the typical d\textit{I}/d\textit{V} point-spectra spectra of the network (orange) as compared to Ag(111) (black). We assign the gap of the $P_5$ assembly to about 0.9-1.0 eV (dashed lines) similar to Ref.\cite{Zhang2020} The spectra also shows a strong resonance at 2.5 V, which we attribute to tunneling into the interface state ($IS$), respectively. d{\it I}/d{\it V} maps (Fig.~\ref{Fig3}b) further reveal the density of states at the valence band at $V_{\rm s}$ = -0.5 V. This atomic feature evolves to a stripe pattern at $V_{\rm s}$ = +2.5 V (Fig.~\ref{Fig3}c), revealing the spatial modulation of the IS state (Fig.~\ref{Fig3}c) similar to the superstructure shown in STM topographic image of Fig.~\ref{Fig2}a. This state, which derived from the occupied Shockley state of the clean Ag(111) surface, is upshifted by more than 2 eV and becomes unoccupied by the presence of the $P_5$ assembly.\cite{GonzalezLakunza2008} 

\noindent
To quantify the local change of work function (LWF), we acquired field-effect resonant tunneling (FERT) spectra  in order to probe IPS between the \textit{cyclo}-$P_{\rm 5}$ assembly and silver.\cite{Borca2010,Borca2020,Liu2021,Liu2022} Experimentally, FERT spectra (also called d\textit{Z}/d\textit{V} spectroscopy) are obtained by sweeping the sample voltage $V_{\rm s}$ while measuring d\textit{I}/d\textit{V} at constant-current by adjusting the tip height $Z$ using the STM feedback loop (see Methods). From a quasi-classical approximation (Fig.~\ref{Fig3}d), tunneling resonances  spectrum occurs when the Fermi level of the tip aligns with the Stark-shifted IPS states and follow the equation :
\begin{equation}
eV_{\rm n} = \phi + \left( \frac{3n\pi\hbar eE}{\sqrt{2m}}\right)
\label{eq1}
\end{equation}
where $V_{\rm n}$ is the sample voltage for the n$^{th}$ IPS, $\phi$ is the work function of the sample, $m$ is the free electron mass and $E$ is the electric field.\\

\noindent
Figure \ref{Fig3}d shows the series of IPS states obtained above the \textit{cyclo}-$P_{\rm 5}$ self-assembly as compared to Ag(111), respectively. The resonance at $V_{\rm s}$ = 2.2 eV of the orange spectrum, which is absent for the Ag one (black), corresponds to the IS  state. The peaks noted $n$ = $1$ to $7$ of the black spectra are the IPS states of the pristine Ag substrate. On the $P_{5}$ assembly, IPS states are clearly shifted to higher voltage when increasing the electric field (i.e. $V_{\rm s}$), which is the signature of the change of LWF.\cite{Liu2021,Liu2022}

\noindent
A quantitative estimation of the LWF can be obtained from Eq.~\ref{eq1}. In Fig.~\ref{Fig3}e,  we plot the voltage position $V_{\rm n}$ of the IPS states as a function of $n^{2/3}$ for both the \textit{cyclo}-$P_{\rm 5}$ network (orange squares) and the Ag(111) substrate (black triangles). By fitting the linear progression of each datasets, we extract the LWF value corresponding to the $y$-intercepts to $\phi_{\rm Ag}$ = 4.49 eV  and $\phi_{\rm P_{\rm 5}}$ = 4.95 eV. Considering that our experimental  estimate of $\phi_{\rm Ag}$ is in excellent agreement with that obtained by ultraviolet photoelectron spectroscopy (UPS),\cite{Hofmann2017} we confirm the strong increase of LWF of $\Delta \phi$ = 0.46 eV induced by the \textit{cyclo}-$P_{\rm 5}$ assembly adsorbed on Ag(111). Altogether, the observation of an IS and the shift of the IPS resonances in tunneling spectroscopy point to a charge transfer from the Ag substrate to the \textit{cyclo}-$P_{\rm 5}$ network and the creation of a strong inwards electric dipole at the interface.

\section*{\large \textsf{Summary and outlook}}
In summary, we synthesized  phosphorus chains and \textit{cyclo}-$P_{\rm 5}$ pentamers by depositing phosphorus atoms on atomically flat Ag(111) in ultra-high vacuum. Using low-temperature AFM with CO-terminated tips, armchair $P$ chains and the planar \textit{cyclo}-$P_{\rm 5}$ rings are resolved with atomic precision. Flat-lying  \textit{cyclo}-$P_{\rm 5}$ pentamers self-assemble into an extended hexagonal assembly in registry with the Ag substrate. DFT calculations support a substantial charge transfer from the Ag substrate to $P_{\rm 5}$ pentamers, which results in a complex charge redistribution at the  $P_{\rm 5}$/Ag interface and the emergence of an interface state. Using force spectroscopic measurements, the inward surface dipole moment induced by this charge transfer is confirmed as an increase of the LCPD value at the $P_{\rm 5}$ assembly in comparison to the pristine Ag.  This corresponds to an increase of LWF at the $P_5$ network as compared to the bare metal substrate. We corroborated these measurement with FERT spectroscopy allowing us to quantify the LWF increase of 0.46 eV at the $P_{\rm 5}$/Ag interface. By exploring the fundamental characteristics of the prototypical \textit{cyclo}-$P_{\rm 5}$/metal interface, our results not only underline the importance of scanning probe microscopy (applicable to other emerging 2D materials and related quantum materials) to study structural and electronic properties at the atomic scale,  but also provides new insights for improved performances of phosphorus-based devices. 
  
\section*{\large \textsf{Methods}}
{\large \textsf{Sample preparation}}\\
The Ag(111) substrate purchased from Mateck GmbH was sputtered by Ar$^{+}$ ions and annealed at 500 °C to eliminate any surface contaminations. Phosphorus atoms were sublimed by heating up a black phosphorus crystal contained in a Knudsen cell in ultra high vacuum (UHV). The P flux was estimated using a quartz microbalance. To obtain the phosphorus chains and $P_{5}$ domains, we annealed the Ag(111) substrate during deposition at temperatures described in Ref.~\cite{Zhang2016}

\noindent
{\large \textsf{STM experiments}}\\
The STM experiments were conducted at a temperature of 4.8 K using an Omicron GmbH low-temperature STM/AFM system operated with Nanonis RC5 electronics.   Differential conductance spectroscopy d{\it I}/d{\it V}(V) spectra were acquired with the lock-in amplifier technique using a modulation of 610 Hz and a modulation amplitude of 10 meV. All voltages refer to the sample bias $V_{\rm s}$ with respect to the tip. For field-emission resonance tunneling spectroscopy (FERT), the lock-in amplifier generates a 15-30 mV (RMS) bias modulation at ~650 Hz. The FERT spectra is obtained by recording the differential conductance data while the sample bias is swept with a closed feedback loop (setpoints: $I_{\rm t}$ = 1 pA, $V_{\rm s}$ = 500 mV).

\noindent
{\large \textsf{AFM experiments}}\\
AFM measurements were performed with commercially available tuning-fork sensors in the qPlus configuration\cite{Giessibl2019} equipped with a tungsten tip ($f_{0}$ = 26 kHz, Q = 10000 to 25 000, nominal spring constant $k$ = 1800, N.m$^{-1}$, oscillation amplitude A $\approx$ 50 pm. Constant-height AFM images were obtained using tips terminated with a single carbon monoxide (CO) in the non-contact mode (frequency-modulated AFM--FMAFM) at zero voltage.\cite{Gross2009,Pawlak2011} CO molecules were adsorbed on the sample maintained at low temperature below 20 K. Before its functionalization, the apex was sharpened by gentle indentations into the silver surface. A single CO molecule was carefully attached to the tip following the procedure of reference.\cite{bartels1998dynamics} Simulations of the AFM images based on the DFT coordinates were carried out using the probe-particle model.\cite{Hapala2014}
Site-dependent $\Delta f(Z)$ spectroscopic measurements to determine the atomic buckling of phosphorus pentamers were obtained with CO-terminated tips. The $\Delta f(V)$ cross-section of 1$\times$85 pixels$^2$ was acquired with Ag-coated metallic tips (tunneling setpoints: $I_{\rm t}$ = 1 pA, $V_{\rm s}$ = 800 mV, $Z_{\rm offset}$ = +80 pm).

\noindent
{\large \textsf{DFT calculations}}\\
All density functional theory calculations were carried out in the Vienna ab initio simulation package (VASP)\cite{Kresse1996} with projector augmented wave (PAW)\cite{Bloechl1994,Kresse1999} method. The generalized gradient approximation (GGA) in the framework of Perdew-Burke-Ernzerhof (PBE)\cite{Perdew1996} was chosen with the plane-wave cutoff energy set at 400 eV for all calculations. The DFT-D3\cite{Grimme2010} method of Grimme was employed to describe the van der Waals (vdW) interactions. The geometries of the structures were relaxed until the force on each atom was less than 0.02 eV \AA$^{-1}$, and the energy convergence criterion of 1$\times$10$^{-4}$ eV was met. The Brillouin zone was sampled using Gamma k-mesh with a separation criterion of 0.03. Metal slabs with 3 atomic layers was adopted as the substrate and the bottom layer was fixed to simulate the bulk. The vacuum spacing between neighboring images was set at least 15 \AA along the non-periodic directions to avoid a periodic interaction. 

\section*{\large Data availability}
The data that supports the findings of this study are available within the paper or its Supplementary Information. All STM/AFM images are raw data. The raw data of spectroscopic measurements are available from the repository ZENODO (link will be updated).

\section*{\large References}
\bibliography{bibliography.bib}

\begin{thebibliography}{10}
\expandafter\ifx\csname url\endcsname\relax
  \def\url#1{\texttt{#1}}\fi
\expandafter\ifx\csname urlprefix\endcsname\relax\def\urlprefix{URL }\fi
\providecommand{\bibinfo}[2]{#2}
\providecommand{\eprint}[2][]{\url{#2}}

\bibitem{Liu2015}
\bibinfo{author}{Liu, H.}, \bibinfo{author}{Du, Y.}, \bibinfo{author}{Deng, Y.}
  \& \bibinfo{author}{Ye, P.~D.}
\newblock \bibinfo{title}{Semiconducting black phosphorus: synthesis, transport
  properties and electronic applications}.
\newblock \emph{\bibinfo{journal}{Chem. Soc. Rev.}}
  \textbf{\bibinfo{volume}{44}}, \bibinfo{pages}{2732--2743}
  (\bibinfo{year}{2015}).

\bibitem{Carvalho2016}
\bibinfo{author}{Carvalho, A.} \emph{et~al.}
\newblock \bibinfo{title}{Phosphorene: from theory to applications}.
\newblock \emph{\bibinfo{journal}{Nat. Rev. Mater.}}
  \textbf{\bibinfo{volume}{1}}, \bibinfo{pages}{1--16} (\bibinfo{year}{2016}).

\bibitem{Batmunkh2016}
\bibinfo{author}{Batmunkh, M.}, \bibinfo{author}{Bat-Erdene, M.} \&
  \bibinfo{author}{Shapter, J.~G.}
\newblock \bibinfo{title}{Phosphorene and phosphorene-based materials –
  prospects for future applications}.
\newblock \emph{\bibinfo{journal}{Adv. Mater.}} \textbf{\bibinfo{volume}{28}},
  \bibinfo{pages}{8586--8617} (\bibinfo{year}{2016}).

\bibitem{Jones1990}
\bibinfo{author}{Jones, R.~O.} \& \bibinfo{author}{Hohl, D.}
\newblock \bibinfo{title}{Structure of phosphorus clusters using simulated
  annealing - {P2} to {P8}}.
\newblock \emph{\bibinfo{journal}{J. Chem. Phys.}}
  \textbf{\bibinfo{volume}{92}}, \bibinfo{pages}{6710--6721}
  (\bibinfo{year}{1990}).

\bibitem{Chen2000}
\bibinfo{author}{Chen, M.~D.}, \bibinfo{author}{Huang, R.~B.},
  \bibinfo{author}{Zheng, L.~S.}, \bibinfo{author}{Zhang, Q.~E.} \&
  \bibinfo{author}{Au, C.~T.}
\newblock \bibinfo{title}{A theoretical study for the isomers of neutral,
  cationic and anionic phosphorus clusters {P5}, {P7}, {P9}}.
\newblock \emph{\bibinfo{journal}{Chem. Phys. Lett.}}
  \textbf{\bibinfo{volume}{325}}, \bibinfo{pages}{22--28}
  (\bibinfo{year}{2000}).

\bibitem{Giusti2021}
\bibinfo{author}{Giusti, L.} \emph{et~al.}
\newblock \bibinfo{title}{Coordination chemistry of elemental phosphorus}.
\newblock \emph{\bibinfo{journal}{Coord. Chem. Rev.}}
  \textbf{\bibinfo{volume}{441}}, \bibinfo{pages}{213927}
  (\bibinfo{year}{2021}).

\bibitem{Cai2010}
\bibinfo{author}{Cai, J.} \emph{et~al.}
\newblock \bibinfo{title}{Atomically precise bottom-up fabrication of graphene
  nanoribbons}.
\newblock \emph{\bibinfo{journal}{Nature}} \textbf{\bibinfo{volume}{466}},
  \bibinfo{pages}{470--473} (\bibinfo{year}{2010}).

\bibitem{Clair2019}
\bibinfo{author}{Clair, S.} \& \bibinfo{author}{de~Oteyza, D.~G.}
\newblock \bibinfo{title}{Controlling a chemical coupling reaction on a
  surface: Tools and strategies for on-surface synthesis}.
\newblock \emph{\bibinfo{journal}{Chem. Rev.}} \textbf{\bibinfo{volume}{119}},
  \bibinfo{pages}{4717--4776} (\bibinfo{year}{2019}).

\bibitem{Zhang2016}
\bibinfo{author}{Zhang, J.~L.} \emph{et~al.}
\newblock \bibinfo{title}{Epitaxial growth of single layer blue phosphorus: a
  new phase of two-dimensional phosphorus}.
\newblock \emph{\bibinfo{journal}{Nano Lett.}} \textbf{\bibinfo{volume}{16}},
  \bibinfo{pages}{4903--4908} (\bibinfo{year}{2016}).

\bibitem{Zhang2021}
\bibinfo{author}{Zhang, W.} \emph{et~al.}
\newblock \bibinfo{title}{Flat epitaxial quasi-{1D} phosphorene chains}.
\newblock \emph{\bibinfo{journal}{Nat. Comm.}} \textbf{\bibinfo{volume}{12}},
  \bibinfo{pages}{5160} (\bibinfo{year}{2021}).

\bibitem{Zhang2020}
\bibinfo{author}{Zhang, W.} \emph{et~al.}
\newblock \bibinfo{title}{Phosphorus pentamers: floating nanoflowers form a
  {2D} network}.
\newblock \emph{\bibinfo{journal}{Adv. Funct. Mater.}}
  \textbf{\bibinfo{volume}{30}}, \bibinfo{pages}{2004531}
  (\bibinfo{year}{2020}).

\bibitem{Yin2022}
\bibinfo{author}{Yin, Y.}, \bibinfo{author}{Gladkikh, V.},
  \bibinfo{author}{Yuan, Q.} \& \bibinfo{author}{Ding, F.}
\newblock \bibinfo{title}{Phosphorus chains and pentamers: The precursors of
  blue phosphorene on the {Ag}(111) substrate}.
\newblock \emph{\bibinfo{journal}{Chem. Mater.}} \textbf{\bibinfo{volume}{34}},
  \bibinfo{pages}{8230--8236} (\bibinfo{year}{2022}).

\bibitem{Li2017}
\bibinfo{author}{Li, L.} \emph{et~al.}
\newblock \bibinfo{title}{Direct observation of the layer-dependent electronic
  structure in phosphorene}.
\newblock \emph{\bibinfo{journal}{Nat. Nanotechnol.}}
  \textbf{\bibinfo{volume}{12}}, \bibinfo{pages}{21--25}
  (\bibinfo{year}{2017}).

\bibitem{Liu2014}
\bibinfo{author}{Liu, H.} \emph{et~al.}
\newblock \bibinfo{title}{Phosphorene: an unexplored {2D} semiconductor with a
  high hole mobility}.
\newblock \emph{\bibinfo{journal}{ACS Nano}} \textbf{\bibinfo{volume}{8}},
  \bibinfo{pages}{4033--4041} (\bibinfo{year}{2014}).

\bibitem{Wang2015}
\bibinfo{author}{Wang, X.} \emph{et~al.}
\newblock \bibinfo{title}{Highly anisotropic and robust excitons in monolayer
  black phosphorus}.
\newblock \emph{\bibinfo{journal}{Nat. Nanotechnol.}}
  \textbf{\bibinfo{volume}{10}}, \bibinfo{pages}{517--521}
  (\bibinfo{year}{2015}).

\bibitem{Shao2014}
\bibinfo{author}{Shao, D.~F.}, \bibinfo{author}{Lu, W.~J.},
  \bibinfo{author}{Lv, H.~Y.} \& \bibinfo{author}{Sun, Y.~P.}
\newblock \bibinfo{title}{Electron-doped phosphorene: {A} potential monolayer
  superconductor}.
\newblock \emph{\bibinfo{journal}{Europhys. Lett.}}
  \textbf{\bibinfo{volume}{108}}, \bibinfo{pages}{67004}
  (\bibinfo{year}{2014}).

\bibitem{Zhang2017}
\bibinfo{author}{Zhang, R.}, \bibinfo{author}{Waters, J.},
  \bibinfo{author}{Geim, A.~K.} \& \bibinfo{author}{Grigorieva, I.~V.}
\newblock \bibinfo{title}{Intercalant-independent transition temperature in
  superconducting black phosphorus}.
\newblock \emph{\bibinfo{journal}{Nat. Comm.}} \textbf{\bibinfo{volume}{8}},
  \bibinfo{pages}{15036} (\bibinfo{year}{2017}).

\bibitem{Li2014a}
\bibinfo{author}{Li, Y.}, \bibinfo{author}{Yang, S.} \& \bibinfo{author}{Li,
  J.}
\newblock \bibinfo{title}{Modulation of the electronic properties of ultrathin
  black phosphorus by strain and electrical field}.
\newblock \emph{\bibinfo{journal}{J. Phys. Chem. C}}
  \textbf{\bibinfo{volume}{118}}, \bibinfo{pages}{23970--23976}
  (\bibinfo{year}{2014}).

\bibitem{Gross2009}
\bibinfo{author}{Gross, L.}, \bibinfo{author}{Mohn, F.}, \bibinfo{author}{Moll,
  N.}, \bibinfo{author}{Liljeroth, P.} \& \bibinfo{author}{Meyer, G.}
\newblock \bibinfo{title}{The chemical structure of a molecule resolved by
  atomic force microscopy}.
\newblock \emph{\bibinfo{journal}{Science}} \textbf{\bibinfo{volume}{325}},
  \bibinfo{pages}{1110--1114} (\bibinfo{year}{2009}).

\bibitem{Moenig2018}
\bibinfo{author}{Mönig, H.} \emph{et~al.}
\newblock \bibinfo{title}{Quantitative assessment of intermolecular
  interactions by atomic force microscopy imaging using copper oxide tips}.
\newblock \emph{\bibinfo{journal}{Nat. Nanotechnol.}}
  \textbf{\bibinfo{volume}{13}}, \bibinfo{pages}{371--375}
  (\bibinfo{year}{2018}).

\bibitem{Kaiser2019}
\bibinfo{author}{Kaiser, K.} \emph{et~al.}
\newblock \bibinfo{title}{An sp-hybridized molecular carbon allotrope,
  cyclo[18]carbon}.
\newblock \emph{\bibinfo{journal}{Science}} \textbf{\bibinfo{volume}{365}},
  \bibinfo{pages}{1299--1301} (\bibinfo{year}{2019}).
\newblock \urlprefix\url{https://doi.org/10.1126/science.aay1914}.

\bibitem{Liu2019}
\bibinfo{author}{Liu, X.} \emph{et~al.}
\newblock \bibinfo{title}{Geometric imaging of borophene polymorphs with
  functionalized probes}.
\newblock \emph{\bibinfo{journal}{Nat. Comm.}} \textbf{\bibinfo{volume}{10}},
  \bibinfo{pages}{1642} (\bibinfo{year}{2019}).

\bibitem{Pawlak2020}
\bibinfo{author}{Pawlak, R.} \emph{et~al.}
\newblock \bibinfo{title}{Quantitative determination of atomic buckling of
  silicene by atomic force microscopy}.
\newblock \emph{\bibinfo{journal}{Proc. Nat. Acad. Sci.}}
  \textbf{\bibinfo{volume}{117}}, \bibinfo{pages}{228--237}
  (\bibinfo{year}{2020}).

\bibitem{Mohn2012}
\bibinfo{author}{Mohn, F.}, \bibinfo{author}{Gross, L.}, \bibinfo{author}{Moll,
  N.} \& \bibinfo{author}{Meyer, G.}
\newblock \bibinfo{title}{Imaging the charge distribution within a single
  molecule}.
\newblock \emph{\bibinfo{journal}{Nat. Nanotechnol.}}
  \textbf{\bibinfo{volume}{7}}, \bibinfo{pages}{227--231}
  (\bibinfo{year}{2012}).

\bibitem{Schuler2014}
\bibinfo{author}{Schuler, B.} \emph{et~al.}
\newblock \bibinfo{title}{Contrast formation in {Kelvin} probe force microscopy
  of single $\pi$-conjugated molecules}.
\newblock \emph{\bibinfo{journal}{Nano Lett.}} \textbf{\bibinfo{volume}{14}},
  \bibinfo{pages}{3342--3346} (\bibinfo{year}{2014}).

\bibitem{Meier2017}
\bibinfo{author}{Meier, T.} \emph{et~al.}
\newblock \bibinfo{title}{Donor–acceptor properties of a single-molecule
  altered by on-surface complex formation}.
\newblock \emph{\bibinfo{journal}{ACS Nano}} \textbf{\bibinfo{volume}{11}},
  \bibinfo{pages}{8413--8420} (\bibinfo{year}{2017}).

\bibitem{Pawlak2017}
\bibinfo{author}{Pawlak, R.} \emph{et~al.}
\newblock \bibinfo{title}{Hydroxyl-induced partial charge states of single
  porphyrins on titania rutile}.
\newblock \emph{\bibinfo{journal}{J. Phys. Chem. C}}
  \textbf{\bibinfo{volume}{121}}, \bibinfo{pages}{3607--3614}
  (\bibinfo{year}{2017}).

\bibitem{Borca2010}
\bibinfo{author}{Borca, B.} \emph{et~al.}
\newblock \bibinfo{title}{Potential energy landscape for hot electrons in
  periodically nanostructured graphene}.
\newblock \emph{\bibinfo{journal}{Phys. Rev. Lett.}}
  \textbf{\bibinfo{volume}{105}}, \bibinfo{pages}{036804}
  (\bibinfo{year}{2010}).

\bibitem{Borca2020}
\bibinfo{author}{Borca, B.} \emph{et~al.}
\newblock \bibinfo{title}{Image potential states of germanene}.
\newblock \emph{\bibinfo{journal}{2D Materials}} \textbf{\bibinfo{volume}{7}},
  \bibinfo{pages}{035021} (\bibinfo{year}{2020}).

\bibitem{Liu2021}
\bibinfo{author}{Liu, X.}, \bibinfo{author}{Wang, L.},
  \bibinfo{author}{Yakobson, B.~I.} \& \bibinfo{author}{Hersam, M.~C.}
\newblock \bibinfo{title}{Nanoscale probing of image-potential states and
  electron transfer doping in borophene polymorphs}.
\newblock \emph{\bibinfo{journal}{Nano Lett.}} \textbf{\bibinfo{volume}{21}},
  \bibinfo{pages}{1169--1174} (\bibinfo{year}{2021}).

\bibitem{Liu2022}
\bibinfo{author}{Liu, X.} \emph{et~al.}
\newblock \bibinfo{title}{Borophene synthesis beyond the single-atomic-layer
  limit}.
\newblock \emph{\bibinfo{journal}{Nat. Mater.}} \textbf{\bibinfo{volume}{21}},
  \bibinfo{pages}{35--40} (\bibinfo{year}{2022}).

\bibitem{Giovanelli2022}
\bibinfo{author}{Giovanelli, L.} \emph{et~al.}
\newblock \bibinfo{title}{On-surface synthesis of unsaturated hydrocarbon
  chains through {C}-{S} activation}.
\newblock \emph{\bibinfo{journal}{Chem. A Eur. Jour.}}
  \textbf{\bibinfo{volume}{28}}, \bibinfo{pages}{e202200809}
  (\bibinfo{year}{2022}).

\bibitem{Boneschanscher2014}
\bibinfo{author}{Boneschanscher, M.~P.}, \bibinfo{author}{Hämäläinen,
  S.~K.}, \bibinfo{author}{Liljeroth, P.} \& \bibinfo{author}{Swart, I.}
\newblock \bibinfo{title}{Sample corrugation affects the apparent bond lengths
  in atomic force microscopy}.
\newblock \emph{\bibinfo{journal}{ACS Nano}} \textbf{\bibinfo{volume}{8}},
  \bibinfo{pages}{3006--3014} (\bibinfo{year}{2014}).

\bibitem{Kawai2018}
\bibinfo{author}{Kawai, S.} \emph{et~al.}
\newblock \bibinfo{title}{Diacetylene linked anthracene oligomers synthesized
  by one-shot homocoupling of trimethylsilyl on {Cu}(111)}.
\newblock \emph{\bibinfo{journal}{ACS Nano}} \textbf{\bibinfo{volume}{12}},
  \bibinfo{pages}{8791--8797} (\bibinfo{year}{2018}).

\bibitem{GonzalezLakunza2008}
\bibinfo{author}{Gonzalez-Lakunza, N.} \emph{et~al.}
\newblock \bibinfo{title}{Formation of dispersive hybrid bands at an
  organic-metal interface}.
\newblock \emph{\bibinfo{journal}{Phys. Rev. Lett.}}
  \textbf{\bibinfo{volume}{100}}, \bibinfo{pages}{156805}
  (\bibinfo{year}{2008}).

\bibitem{Nony2009}
\bibinfo{author}{Nony, L.}, \bibinfo{author}{Foster, A.~S.},
  \bibinfo{author}{Bocquet, F.} \& \bibinfo{author}{Loppacher, C.}
\newblock \bibinfo{title}{Understanding the atomic-scale contrast in {Kelvin}
  probe force microscopy}.
\newblock \emph{\bibinfo{journal}{Phys. Rev. Lett.}}
  \textbf{\bibinfo{volume}{103}}, \bibinfo{pages}{036802}
  (\bibinfo{year}{2009}).

\bibitem{Sadeghi2012}
\bibinfo{author}{Sadeghi, A.} \emph{et~al.}
\newblock \bibinfo{title}{Multiscale approach for simulations of {Kelvin} probe
  force microscopy with atomic resolution}.
\newblock \emph{\bibinfo{journal}{Phys. Rev. B}} \textbf{\bibinfo{volume}{86}},
  \bibinfo{pages}{075407} (\bibinfo{year}{2012}).

\bibitem{Hofmann2017}
\bibinfo{author}{Hofmann, O.~T.} \emph{et~al.}
\newblock \bibinfo{title}{Orientation-dependent work-function modification
  using substituted pyrene-based acceptors}.
\newblock \emph{\bibinfo{journal}{J. Phys. Chem. C}}
  \textbf{\bibinfo{volume}{121}}, \bibinfo{pages}{24657--24668}
  (\bibinfo{year}{2017}).

\bibitem{Giessibl2019}
\bibinfo{author}{Giessibl, F.~J.}
\newblock \bibinfo{title}{The {qPlus} sensor, a powerful core for the atomic
  force microscope}.
\newblock \emph{\bibinfo{journal}{Rev. Sci. Instrum.}}
  \textbf{\bibinfo{volume}{90}}, \bibinfo{pages}{011101}
  (\bibinfo{year}{2019}).

\bibitem{Pawlak2011}
\bibinfo{author}{Pawlak, R.}, \bibinfo{author}{Kawai, S.},
  \bibinfo{author}{Fremy, S.}, \bibinfo{author}{Glatzel, T.} \&
  \bibinfo{author}{Meyer, E.}
\newblock \bibinfo{title}{Atomic-scale mechanical properties of orientated
  {C}$_{60}$ molecules revealed by noncontact atomic force microscopy}.
\newblock \emph{\bibinfo{journal}{ACS Nano}} \textbf{\bibinfo{volume}{5}},
  \bibinfo{pages}{6349--6354} (\bibinfo{year}{2011}).

\bibitem{bartels1998dynamics}
\bibinfo{author}{Bartels, L.} \emph{et~al.}
\newblock \bibinfo{title}{Dynamics of electron-induced manipulation of
  individual {CO} molecules on {C}u (111)}.
\newblock \emph{\bibinfo{journal}{Phys. Rev. Lett.}}
  \textbf{\bibinfo{volume}{80}}, \bibinfo{pages}{2004} (\bibinfo{year}{1998}).

\bibitem{Hapala2014}
\bibinfo{author}{Hapala, P.} \emph{et~al.}
\newblock \bibinfo{title}{Mechanism of high-resolution {STM/AFM} imaging with
  functionalized tips}.
\newblock \emph{\bibinfo{journal}{Phys. Rev. B}} \textbf{\bibinfo{volume}{90}},
  \bibinfo{pages}{085421} (\bibinfo{year}{2014}).

\bibitem{Kresse1996}
\bibinfo{author}{Kresse, G.} \& \bibinfo{author}{Furthmüller, J.}
\newblock \bibinfo{title}{Efficient iterative schemes for ab initio
  total-energy calculations using a plane-wave basis set}.
\newblock \emph{\bibinfo{journal}{Phys. Rev. B}} \textbf{\bibinfo{volume}{54}},
  \bibinfo{pages}{11169--11186} (\bibinfo{year}{1996}).

\bibitem{Bloechl1994}
\bibinfo{author}{Blöchl, P.~E.}
\newblock \bibinfo{title}{Projector augmented-wave method}.
\newblock \emph{\bibinfo{journal}{Phys. Rev. B}} \textbf{\bibinfo{volume}{50}},
  \bibinfo{pages}{17953--17979} (\bibinfo{year}{1994}).

\bibitem{Kresse1999}
\bibinfo{author}{Kresse, G.} \& \bibinfo{author}{Joubert, D.}
\newblock \bibinfo{title}{From ultrasoft pseudopotentials to the projector
  augmented-wave method}.
\newblock \emph{\bibinfo{journal}{Phys. Rev. B}} \textbf{\bibinfo{volume}{59}},
  \bibinfo{pages}{1758--1775} (\bibinfo{year}{1999}).

\bibitem{Perdew1996}
\bibinfo{author}{Perdew, J.~P.}, \bibinfo{author}{Burke, K.} \&
  \bibinfo{author}{Ernzerhof, M.}
\newblock \bibinfo{title}{Generalized gradient approximation made simple}.
\newblock \emph{\bibinfo{journal}{Phys. Rev. Lett.}}
  \textbf{\bibinfo{volume}{77}}, \bibinfo{pages}{3865--3868}
  (\bibinfo{year}{1996}).

\bibitem{Grimme2010}
\bibinfo{author}{Grimme, S.}, \bibinfo{author}{Antony, J.},
  \bibinfo{author}{Ehrlich, S.} \& \bibinfo{author}{Krieg, H.}
\newblock \bibinfo{title}{A consistent and accurate ab initio parametrization
  of density functional dispersion correction ({DFT}-{D}) for the 94 elements
  {H}-{Pu}}.
\newblock \emph{\bibinfo{journal}{J. Chem. Phys.}}
  \textbf{\bibinfo{volume}{132}}, \bibinfo{pages}{154104}
  (\bibinfo{year}{2010}).

\end{thebibliography}

\section*{\large Acknowledgments}
E.M. and R.P. acknowledge funding from the Swiss Nanoscience Institute (SNI), the European Research Council (ERC) under the European Union’s Horizon 2020 research and innovation programme (ULTRADISS grant agreement No 834402 and supports as a part of NCCR SPIN, a National Centre of Competence (or Excellence) in Research, funded by the SNF (grant number 51NF40-180604).
E.M., T.G. and S.-X.L. acknowledge the Sinergia Project funded by the SNF (CRSII5\_213533). E.M., T.G. and R.P. acknowledge the SNF grant (200020\_188445). 
T.G. acknowledges the FET-Open program (Q-AFM grant agreement No 828966) of the European Commission. J.-C.L. acknowledges funding from the European Union’s Horizon 2020 research and innovation programme under the Marie Sklodowska-Curie grant agreement number 847471.
C.L. and E.M. acknowledges the Georg H. Endress Foundation.		
		
\section*{\large Author information}
{\large \textsf{Authors and Affiliations}}\\
\noindent
{\bf Department of Physics, University of Basel, Klingelbergstrasse 82, 4056 Basel, Switzerland}\\
Outhmane Chahib, Chao Li, Jung-Ching Liu, Thilo Glatzel, Ernst Meyer \& Rémy Pawlak

\noindent
{\bf State Key Laboratory of Precision Spectroscopy School of Physics and Electronic Science,
East China Normal University, 500 Dongchuan Road, Shanghai 200241, China}\\
Yulin Yin \& Qinghong Yuan,

\noindent
{\bf Faculty of Materials Science and Engineering/Institute of Technology for Carbon Neutrality, Shenzhen Institute of Advanced Technology, Chinese Academy of Sciences, Shenzhen 518055, China}\\
Feng Ding


\noindent
{\large \textsf{Contributions}}\\
R.P. and E.M. conceived the experiments. O.C. and R.P. performed the STM/AFM measurements. Y.Y. F.D. and Q.Y. performed DFT calculations. O.C. and R.P. analyzed the data. R.P. wrote the manuscript. All authors discussed on the results and revised the manuscript.

 \noindent
{\large \textsf{Corresponding authors
}}\\
Correspondence to ernst.meyer@unibas.ch or remy.pawlak@unibas.ch

\section*{\large Ethics declarations}
\noindent
{\large \textsf{Competing interests}}\\
The authors declare no competing interests.


\newpage

\end{document}